\documentclass[letterpaper, 10 pt, conference]{ieeeconf}  

\IEEEoverridecommandlockouts                              

\usepackage{amsmath} 
\usepackage{amssymb}  
\usepackage{mathtools}
\usepackage{subcaption,tikz}
\usepackage{caption}
\usepackage[font=small]{caption}
\newtheorem{definition}{Definition}
\newtheorem{theorem}{Theorem}
\newtheorem{assumption}{Assumption}
\newtheorem{remark}{Remark}
\DeclareMathOperator*{\minimize}{minimize}
\DeclareMathOperator*{\subject_to}{s. t. }

\title{\LARGE \bf
Data-Based Moving Horizon Estimation for Linear Discrete-Time Systems
}

\author{Tobias M. Wolff$^{1}$, Victor G. Lopez$^{1}$ and Matthias A. Müller$^{1}$ 
\thanks{*This project has received funding from the European Research Council (ERC) under the European Union’s Horizon 2020 research and innovation programme (grant agreement No 948679).}
\thanks{$^{1}$Leibniz University Hannover, Institute of Automatic Control, 30167
Hannover, Germany. E-mail:
        {\tt\small \{wolff, lopez,mueller\}@irt.uni-hannover.de}}%
}

\newcommand\copyrighttext{%
	\footnotesize \copyright 2022 IEEE. Personal use of this material is permitted. Permission from IEEE must be obtained for all other uses, in any current or future media, including reprinting/republishing this material for advertising or promotional purposes, creating new collective works, for resale or redistribution to servers or lists, or reuse of any copyrighted component of this work in other works.}
\newcommand\copyrightnotice{%
	\begin{tikzpicture}[remember picture,overlay]
		\node[anchor=south,yshift=10pt] at (current page.south) {\fbox{\parbox{\dimexpr\textwidth-\fboxsep-\fboxrule\relax}{\copyrighttext}}};
	\end{tikzpicture}%
}

\begin{document}

\maketitle
\thispagestyle{empty}
\pagestyle{empty}
\copyrightnotice

\begin{abstract}
This paper introduces a data-based moving horizon estimation (MHE) scheme for linear time-invariant discrete-time systems. The scheme solely relies on collected data without employing any system identification step. Robust global exponential stability of the data-based MHE is proven under standard assumptions for the case where the online output measurements are corrupted by some non-vanishing measurement noise. A simulation example illustrates the behavior of the data-based MHE scheme.
\end{abstract}
\section{INTRODUCTION}
State estimation is indispensable for many applications such as control, monitoring, or fault diagnosis. For linear systems, the Kalman Filter gives the optimal estimate when the process and the measurement noise are white, normally distributed processes \cite{Kalman1960}. An alternative that performs particularly well for nonlinear systems is moving horizon estimation (MHE) which, additionally, offers the possibility of incorporating constraints on estimated systems states \cite{Rawlings2017}.

The standard design of an MHE requires that a mathematical model of the underlying system is available. This mathematical model can be derived by exploiting identification strategies like subspace identification algorithms that are based on available input/output data to determine the system's parameters \cite{Van1997}. However, due to the presence of noise and the complexity of real systems, an identified model often contains parametric inaccuracies and unmodeled dynamics. 

In \cite{Willems2005}, it was shown that a single persistently exciting trajectory of a controllable linear time-invariant discrete-time (LTI-DT) system spans all trajectories of that system. This result, also commonly called Willems' fundamental lemma, has recently been used for the design of controllers purely based on input/output and possibly state data of the system, compare, e.g., the results of \cite{Coulson2019,De2019,Berberich2020_MPC, Lopez2021} as well as the recent survey \cite{Markovsky2021} and the references therein. Note that all of these results develop direct data-based controllers, i.e., no intermediate step is taken to identify a system model. 

These considerations naturally lead to the question whether it is also possible to design estimators purely based on collected persistently exciting input/output data from the system. To this end, much fewer results are available in the literature compared to data-based control. To the best of the authors' knowledge, the first work in this context seems to come from the field of fault diagnosis without exploiting the fundamental lemma \cite{Ding2011}. Very recently, two results appeared which exploit the fundamental lemma. First, in \cite{Turan2021}, the design of a data-driven unknown-input observer was proposed. Second, the design of an observer based on the duality principle of control and estimation was presented in \cite{Adachi2021}. 

One core problem of data-based estimation, which is well known from the field of subspace identification, is that a state sequence can only be determined up to an unknown similarity transformation when only input/output data, but no other system knowledge, is available \cite{Van1997}. This result even holds true for the nominal case, in which no disturbances affect the input/output data. As a consequence, such an estimated state sequence will most likely not have any physical meaning. This might not be a problem if the estimated system states are only needed for control purposes. However, in monitoring applications one is usually interested in physically meaningful state estimates. One possibility to overcome this problem is to incorporate system knowledge in form of an offline measured state sequence as in \cite{Turan2021} and \cite{Adachi2021}. This form of additional system knowledge is also exploited in this work. 

The contribution of this paper is two-fold. First, we present a purely data-based MHE scheme, which, to the best of the authors' knowledge, has not been introduced in the literature so far. Additionally, we propose how measurement noise in the online phase can be incorporated in this scheme. Second, we prove robust global exponential stability (RGES) of the data-based MHE for linear detectable systems and quadratic cost functions.

In Section \ref{sec:preliminaries}, we state some technical definitions and the setup. The core contributions of this paper are given in Section \ref{sec:Robust_MHE}, in which we focus on the robust MHE scheme that considers measurement noise and for which we show RGES. We close this paper with an illustrative example and a conclusion in Sections \ref{sec:Illustrative_Example} and \ref{sec:Conclusion}, respectively.

\section{Preliminaries and Setup}
\label{sec:preliminaries}
The set of integers in the interval $[a,b] \subseteq \mathbb{R}$ is denoted by $\mathbb{I}_{[a,b]}$ and the set of integers greater than or equal to $a$ by $\mathbb{I}_{\geq a}$. For a vector $x = [x_1 \dots x_n]^\top \in \mathbb{R}^n$ and a symmetric positive definite matrix $P$, we write $|x|_P = \sqrt{x^\top P x} $. The identity matrix of dimension $n$ is denoted by $I_n$. The Euclidean norm $||x||_2$ is written as $|x|$. A stacked window of a sequence $\{x(k) \}_{k=0}^{N-1}$ is written as $x_{[0,N-1]} = \begin{bmatrix} x(0)^\top & \hdots & x(N-1)^\top  \end{bmatrix}^\top$.

A function $\alpha : \mathbb{R}_{\geq 0} \rightarrow \mathbb{R}_{\geq 0} $ is of class $\mathcal{K}_\infty$, if $\alpha$ is continuous, strictly increasing, unbounded and $\alpha(0) = 0$. The notation $\lceil a \rceil $ for $a \in  \mathbb{R}$ is defined as the smallest integer greater than or equal to $a$. A pairwise maximization is denoted by $\mathrm{max}(a,b)= a \oplus b$ for $a$, $b \in \mathbb{R}$. Furthermore, let
\begin{equation*}
\bigoplus_{k = t_1}^{t_2} a_k\coloneqq \max_ {t_1 \leq k \leq t_2} a_k
\end{equation*}
for all $a_k \in \mathbb{R}$ and for some $t_1$, $t_2 \in \mathbb{I}_{\geq 0}$.

The Hankel matrix of depth $L$ of a stacked window ${x}_{[0,N-1]}$ is defined by
\begin{equation*}
H_L({x}_{[0,N-1]}) = \begin{pmatrix}
x(0) & x(1) & \dots & x(N-L) \\
x(1) & x(2) & \dots & x(N-L+1)\\
\vdots & \vdots & \ddots & \vdots \\
x(L-1) & x(L) & \dots & x(N-1) \\
\end{pmatrix} \hspace{-4pt}.
\end{equation*}

We recapitulate briefly the fundamental lemma stated in~\cite{Willems2005}. It holds for LTI-DT systems of the form
\begin{subequations} \label{system}
\begin{align}
x(t+1) &= Ax(t) +Bu(t)\\
y(t) &= Cx(t) + Du(t),
\end{align}
\end{subequations}
where $u(t) \in \mathbb{R}^m$, $x(t) \in \mathbb{R}^n$, and $y(t) \in \mathbb{R}^p$. The statement of the fundamental lemma includes the notion of persistency of excitation.
\begin{definition}
An input sequence $\{u(k)\}_{k=0}^{N-1}$ is persistently exciting of order $L$ if $\mathrm{rank}(H_L(u_{[0,N-1]})) = mL$.
\end{definition}
Originally, the fundamental lemma is based on behavioral systems theory. Its formulation in the classical state space framework is presented in \cite{Berberich2020_trajectory,VanWarde2020} as follows.
\begin{theorem}
	\label{thm:Willems}
(\cite{Berberich2020_trajectory}) Suppose $u_{[0,N-1]}$, $y_{[0,N-1]}$ is a trajectory of a controllable LTI system (\ref{system}), where $u_{[0,N-1]}$ is persistently exciting of order $L+n$. Then, $\overline{u}_{[0,L-1]}$, $\overline{y}_{[0,L-1]}$ is a trajectory of system (\ref{system}) if and only if there exists $\alpha \in \mathbb{R}^{N-L +1}$ such that
\begin{equation}
\begin{bmatrix}
H_L(u_{[0,N-1]}) \\
H_L(y_{[0,N-1]}) \\
\end{bmatrix} \alpha = \begin{bmatrix}
\overline{u}_{[0,L-1]}\\
\overline{y}_{[0,L-1]} \\
\end{bmatrix} \label{Willems_Lemma} .
\end{equation}
\end{theorem}
\vspace{0.2cm}
Loosely speaking, the interpretation of this fundamental lemma is that any length $L$ input/output trajectory of an LTI-DT system is contained within the span of length $L$ windows of one single persistently exciting trajectory. 

\begin{remark}
\label{Minimal_realization}
In addition to (\ref{Willems_Lemma}), it holds that \cite[Eq. (5)]{Berberich2020_trajectory}
\begin{equation}
\overline{x}_{[0, L-1]} = \sum_{i = 0}^{N-L} \alpha_i x_{[i, L-1 +i]},
\end{equation}
where $\overline{x}_{[0, L-1]}$ and $x_{[0, L-1]}$ are state trajectories of system~(\ref{system}) that correspond to the trajectories $\overline{u}_{[0,L-1]}$, $\overline{y}_{[0,L-1]}$ and $u_{[0,N-1]}$, $y_{[0,N-1]}$ of (\ref{Willems_Lemma}), respectively. The state trajectories $\overline{x}_{[0, L-1]}$ and $x_{[0, L-1]}$ need not be unique since we do not assume that system (\ref{system}) is necessarily in a minimal realization.
\end{remark}

In this paper, we consider an LTI-DT system of the form 
\begin{subequations} \label{system_def}
\begin{align}
x(t+1) &= Ax(t) +Bu(t) \label{state_dynamics}\\
\tilde{y}(t) &= Cx(t) + Du(t) + v(t) 
\end{align}
\end{subequations}
that differs from (\ref{system}) in the presence of some non-vanishing measurement noise $v(t) \in \mathbb{V}$. We use the notation $y$ instead of $\tilde{y}$ whenever $v = 0$.

The setup of this paper considers two main phases: an offline and an online phase. In the offline phase, we assume having noise-free measurements of the inputs, outputs, and states. This assumption was also taken in \cite{Turan2021,Adachi2021}. Although this assumption can be rather restrictive in general, it is, e.g., satisfied in applications in which the states of a manufactured product can be measured in a dedicated laboratory using sophisticated and potentially expensive measurement hardware.

This assumption is crucial for our data-based MHE scheme. In the estimation process, the available offline data is combined with the measured online data to compute $\alpha$. This $\alpha$ together with the offline collected state sequence determines the estimated state sequence. As mentioned in Remark \ref{Minimal_realization}, this estimated state sequence corresponds to the same realization (\ref{system_def}) as the state sequence of the offline collected state sequence. 

In the online phase, the system states are no longer measurable and need to be estimated. Moreover, the online output measurements are corrupted by some noise $v \in \mathbb{V}$. 

Throughout this paper, we assume that system (\ref{system_def}) is controllable and that the offline collected input sequence $\{ u^d(k)\} _{k=0}^{N-1}$ is persistently exciting of order $L+n$, where $L$ denotes the horizon length of the data-based MHE scheme.

\section{Robust Data-Based MHE}
\label{sec:Robust_MHE}
In this section, we introduce the robust data-based MHE scheme: at each time $t > L -1$, solve
\begin{subequations} \label{MHE_noisy}
\begin{align}
&\minimize_{\overline{x}_{[-L+1,0]}(t), \alpha(t), \sigma_{[-L+1,0]}^y (t)} \hspace{0.15cm}  J_L(\overline{x}(t-L+1|t), \sigma^y_{[-L+1,0]}(t)) \label{cost_function_noisy} \\
&\subject_to \hspace{0.2cm} \begin{bmatrix}
H_L(u^d_{[0,N-1]}) \\
H_L(y^d_{[0,N-1]}) \\
H_L(x^d_{[0,N-1]}) \\
\end{bmatrix}
\alpha(t) =
\begin{bmatrix}
u_{[t-L+1,t]}\\
\tilde{y}_{[t-L+1,t]} -\sigma_{[-L+1,0]}^y (t)\\
\overline{x}_{[-L + 1,0]}(t) \\
\end{bmatrix} \label{System_Dynamics_noisy}\\
& \hspace{2cm} \overline{x}_{[-L+1,0]}(t) \in \mathcal{X} \label{state_constraint_noisy}
\end{align}
where
\begin{align}
J_L(\overline{x}&(t-L+1|t), \sigma^y_{[-L+1,0]}(t)) \nonumber \\
=& \rho \Gamma(\overline{x}(t-L+1|t)) +   \sum_{k= t-L +1}^{t}  \ell(\sigma^y(k|t)) \label{MHE_noisy_cost}
\end{align}
\end{subequations}
and $\rho >0$, which is needed to prove RGES later on. A few comments are in order to explain the scheme in detail. The online measured input and output sequences are denoted by $u_{[t-L+1,t]}$ and $\tilde{y}_{[t-L+1,t]}$. To distinguish measured variables from decision variables, we write the time $t$ in the index for measured variables, e.g., $\tilde{y}_{[t-L+1,t]}$. The sequences $u^d_{[0,N-1]}$, $y^d_{[0,N-1]}$ and $x^d_{[0,N-1]}$ are the offline collected data. Furthermore, $\bar{x}_{[-L + 1,0]}(t) \coloneqq [\bar{x}(t-L+1|t)^\top, \dots, \bar{x}(t|t)^\top ]^\top$ is one of the decision variables of problem (\ref{MHE_noisy}) and denotes the estimated state sequence from time $t-L+1$ up to time $t$, estimated at time $t$. We denote the optimizers of problem (\ref{MHE_noisy}) by $\hat{x}_{[-L + 1,0]}(t)$, $\hat{\sigma}_{[-L+1,0]}(t)$, and $\hat{\alpha}(t)$. The state estimate at each time $t$ is then defined as $\hat{x}(t):=\hat{x}(t|t)$. The set $\mathcal{X} \subseteq \mathbb{R}^n$ denotes the state constraint set. As usual in MHE \cite{Mueller2017,Allan2019}, these constraints can be included in (\ref{MHE_noisy}) in order to improve the estimation, if it is known that they are inherently satisfied by the system states (such as, e.g., nonnegativity constraints for concentrations in a chemical reactor). We introduce the fitting error $\sigma^y$ on the right-hand side of (\ref{System_Dynamics_noisy}). Without this fitting error, the vector $\alpha$ fulfilling the second row of (\ref{System_Dynamics_noisy}) may not necessarily exist, because the output measurements that are corrupted by noise are not necessarily in the span of the noise-free measurements of the output. Similar to model-based MHE formulations \cite{Mueller2017,Allan2019}, the fitting error $\sigma^y$ is penalized in the cost function, which hence trades off how much we believe our prior and how much we believe our measurements. The prior weighting $\Gamma$ at time instant $t$ penalizes the difference between the estimated state at the beginning of the horizon, $\bar{x}(t-L+1|t)$, and the prior estimate\footnote{This prior is typically called filtering prior in the MHE literature, compare, e.g., \cite{Allan2019}.} $\hat{x}(t-L+1)$ corresponding to the state estimate at time $t-L+1$, i.e.,
\begin{equation}
\Gamma(\overline{x}(t-L+1|t)) = |\overline{x}(t-L+1|t) - \hat{x}(t-L+1)|_P^2 \label{eq:Gamma_MHE}
\end{equation}
for some preselected positive definite matrix $P$. By means of this prior weighting, one takes into account the previous measurements that are not part of the current horizon, compare the detailed discussion in \cite{Rao2001}. 

Finally, we note the following subtle difference compared to various existing stability proofs for model-based MHE schemes. While these (for simplicity) often take into account only output measurements up to time $t-1$ for estimating $x(t)$ (compare, e.g.,  \cite{Mueller2017,Rawlings2017,Allan2019}), we also use the current measurement $y(t)$ in (\ref{System_Dynamics_noisy}) in order to determine $\alpha(t)$.

When not enough measurements are available yet to fill a complete horizon $L$, the data-based scheme is slightly adapted. Namely, for $ t \leq L-1$, solve
\begin{subequations}
\label{MHE_nom_ini}
\begin{align}
&\minimize_{\overline{x}_{[0,t]}(t), \alpha(t), \sigma^y_{[0,t]}} \hspace{0.2cm}   J_{t+1}(\overline{x}(0|t), \sigma^y_{[0,t]}(t))\label{cost_function_nom_ini} \\
&\subject_to \hspace{0.2cm} \begin{bmatrix}
H_{t+1}(u^d_{[0,N-1]}) \\
H_{t+1}(y^d_{[0,N-1]}) \\
H_{t+1}(x^d_{[0,N-1]}) \\
\end{bmatrix}
\alpha(t) =
\begin{bmatrix}
u_{[0,t]}\\
\tilde{y}_{[0,t]} - \sigma^y_{[0,t]}\\
\overline{x}_{[0,t]}(t) \\
\end{bmatrix} \label{System_Dynamics_nom_ini}\\
& \hspace{2cm} \overline{x}_{[0,t]}(t) \in \mathcal{X} \label{state_constraint_nom_ini}
\end{align}
where
\begin{align}
J_{t+1} (\overline{x}(0|t),\sigma^y_{[0,t]}(t)) & = \rho \Gamma(\overline{x}(0|t)) +\sum_{k= 0}^{t}  \ell(\sigma^y(k|t))   
\end{align}
\end{subequations}
and
\begin{equation}
\Gamma(\overline{x}(0|t)) = |\overline{x}(0|t) - \hat{x}_0|_P^2 \label{eq:Gamma_initial}
\end{equation}
for a prior estimate $\hat{x}_0$ (of the real, unknown initial condition $x_0$). In this initial time interval, the dimensions of the Hankel matrices change at every iteration, because the length of the estimated trajectory is different for every time instant.

\begin{remark}
Similar to model-based MHE, this initial phase is treated here by using the full information estimator (FIE), meaning that all available measurements are taken into account to estimate the system's states \cite{Rawlings2017}. Its application to the case $t > L-1$, in general, becomes intractable at some point, because the complexity of the optimization problem grows continuously. Obviously, this problem also holds true for the data-based case. Additionally, a similar data-based FIE would require the collection of an infinite, persistently exciting input/output trajectory, which is impossible. 
\end{remark}

\begin{remark}
In contrast to recently proposed data-based model predictive control (MPC) schemes, no regularization term for $\alpha$ is included in the cost function, compare \cite[Eq. (6)]{Coulson2019,Berberich2020_MPC}.
This is due to our assumption that the offline collected data is noise-free and, therefore, there is no risk of $\alpha$ amplifying the (offline) noise sequence.
\end{remark}

In the following, we prove RGES of the presented data-based MHE scheme (\ref{MHE_noisy}) and (\ref{MHE_nom_ini}). To this end, we rely on a proof technique that was originally developed to show robust asymptotic stability (RAS) of a nonlinear model-based MHE~\cite{Mueller2017} and that was recently streamlined in \cite{Allan2019}. Here, we suitably extend it to the data-based case, while exploiting simplifications induced by linearity of the (unknown) system dynamics\footnote{Earlier model-based MHE proofs that were specifically developed for linear systems have mostly been done under the assumption that no disturbances are present \cite{Rao2001} or result in similar estimation error bounds that get worse with increasing horizon $L$ \cite{Alessandri2014}, similar to \cite{Mueller2017,Allan2019}.}.

From here on, we denote sequences of finite or infinite length by bold face symbols $\mathbf{v} \coloneqq \{ v(t_1), \dots, v(t_2)\}$ for some $t_1$, $t_2 \in \mathbb{I}_{\geq 0}$ or $\mathbf{v} \coloneqq \{ v(t_1), v(t_2), \dots\}$, respectively, to simplify the notation. The solution to (\ref{system_def}) at time $t$ for initial condition $x_0$ and input sequence $\mathbf{u} = \{ u(0), u(1), \dots \}$ is denoted by $x(t; x_0, \mathbf{u})$\footnote{With slight abuse of notation, we denote by $\mathbf{x}(x,\mathbf{u})$ and $h(\mathbf{x}, \mathbf{u})$ the state and corresponding output sequence starting at initial condition $x$ and generated by the input sequence $\mathbf{u}$.}. Additionally, let $||\mathbf{v}|| \coloneqq \mathrm{sup}_{t \geq 0}|v(t)|$ denote the supremum norm of sequence $\mathbf{v}$ and $||\mathbf{v}||_ {[a,b]} \coloneqq \mathrm{sup}_{t \in [a,b]}|v(t)|$. We assume having a linear detectable system. A linear detectable system implies the property of ``incremental exponential uniform output-to-state stability" (e-UOSS) (compare \cite[Theorem 6]{Knufer2020}, \cite{Rawlings2017}). 
\begin{definition}
\label{Definition_UOSS_Text}
The system $ x(t+1) =f(x(t), u(t))$ and $y(t) =h(x(t), u(t)) $ is e-UOSS if there exist constants  $c_\beta \geq~1$, $\lambda_\beta \in (0,1)$ and a function $\gamma_d \in \mathcal{K}_\infty$ such that for each pair of initial conditions $x_1$, $x_2 \in \mathbb{R}^n$ and any input sequence $\mathbf{u} $ generating the state sequences $\mathbf{x}_1(x_1, \mathbf{u})$ and $\mathbf{x}_2(x_2, \mathbf{u})$ the following holds for all $\tau \in \mathbb{I}_{\geq 0}$: 
\begin{align}
|x(\tau; x_1, \mathbf{u}) &- x(\tau; x_2,\mathbf{u})| \leq  c_\beta|x_1 -x_2| \lambda_\beta^\tau \nonumber \\  &\oplus  \gamma_d(||h(\mathbf{x}_1, \mathbf{u}) -h(\mathbf{x}_2, \mathbf{u})|| _{[0, \tau-1]}).  \label{Definition_UOSS}
\end{align}
\end{definition}
The term ``uniform" is employed, since the right-hand side of (\ref{Definition_UOSS}) holds uniformly for all $\mathbf{u}$ (with the same $c_{\beta}$, $\lambda_{\beta}$ and $\gamma_d$) \cite{Rawlings2017}. Furthermore, since we consider a linear detectable system, the function $\gamma_d$ can be chosen as a linear function $\gamma_d(r)  =  c_{\gamma_d} r$ without loss of generality \cite[Theorem 6]{Knufer2020}. In the following, we will not use the general $\mathcal{K}_\infty$ function but the linear function when referring to the e-UOSS property.

Usually, in the case of nonlinear model-based MHE, ``incremental input output-to-state stability" (i-IOSS) is considered when RAS is proven \cite{Mueller2017, Allan2019}. In the linear case that is considered here, we use (i) an exponential version which is implied by detectability of $(A, C)$ (compare, e.g., \cite{Knufer2020}) and (ii) only consider an output in~(\ref{Definition_UOSS}) and not an additional (disturbance) input, since we only consider measurement noise in (\ref{system_def}) but not process noise (i.e., disturbances in the dynamics (\ref{state_dynamics})). An extension to the latter setting is subject of ongoing research. In order to prove RGES, we consider the following assumptions:
\begin{assumption}
\label{ass:detectability}
The pair ($A$,$C$) of system (\ref{system_def}) is detectable. 
\end{assumption}

\begin{assumption}
\label{ass:quadrati_costs}
The stage costs $\ell$ and the prior weighting $\Gamma$ in (\ref{MHE_noisy}) and (\ref{MHE_nom_ini}) are positive definite and quadratic. 
\end{assumption}
In order to fulfill Assumption \ref{ass:quadrati_costs}, one can choose $\ell (s) = |s|^2_R$ (with $R$ positive definite) and $\Gamma$ as defined in (\ref{eq:Gamma_MHE}) and~(\ref{eq:Gamma_initial}). Assumption \ref{ass:quadrati_costs} implies that the stage costs and the prior weighting can be lower and upper bounded as follows
\begin{align*}
p_1 s^2 \leq  \Gamma (s)& \leq  p_2 s^2, \\
r_1 s^2 \leq  \ell(s)& \leq r_2 s^2,
\end{align*}
where $p_1 \coloneqq  \lambda_{\mathrm{min}}(P)$, $p_2 \coloneqq  \lambda_{\mathrm{max}}(P)$ and $ r_1 \coloneqq \lambda_{\mathrm{min}}(R)$, $r_2 \coloneqq \lambda_{\mathrm{max}}(R)$.
We use the following definition of RGES. 
\begin{definition}
\label{def:GAS}
Consider system (\ref{system_def}), subject to disturbance $v \in \mathbb{V}$. A state estimator is RGES if there exist constants $c_{1}$, $c_{2} \geq 1$ and $\lambda_{1}$, $\lambda_{2} \in (0,1)$ such that for all $x_0$, $\hat{x}_0 \in \mathbb{R}^n$, all $v \in \mathbb{V}$ the following is satisfied for all $t \in \mathbb{I}_{\geq 0}$: 
\begin{align}
|x(t) - \hat{x}(t)| \leq  &c_{1} |x_0 - \hat{x}_0| \lambda_{1}^t \nonumber \\
& \oplus \: \bigoplus_{\tau = 0}^{t} c_{2} |v(t-\tau)| \lambda_{2}^\tau .  \label{eq:RGES}
\end{align}
\end{definition}
Note that such an RGES definition using also discounting of the disturbance terms has been introduced in the model-based MHE context in \cite[Definition 1 and 2]{Knufer2018} and has (sometimes in a similar asymptotic form) also been used in various recent MHE results, compare, e.g., \cite{Knuefer2021,Allan2020,Schiller2021}. 

RGES of the proposed data-based MHE is shown in the following theorem.
\begin{theorem}
\label{thm:RGES}
\textit{(RGES of MHE)} Suppose Assumptions~\ref{ass:detectability}~-~\ref{ass:quadrati_costs} hold. Then, there exist $T$ and $\rho$ such that if $L \geq T$, the MHE scheme defined in (\ref{MHE_noisy}) and (\ref{MHE_nom_ini}) is RGES, i.e., there exist $c_1$, $c_2\geq 1$ and $\lambda_{1}$, $\lambda_{2} \in (0,1)$ such that for all $x_0$, $\hat{x}_0 \in \mathbb{R}^n$ and for all $t \in \mathbb{I}_{\geq 0}$ (\ref{eq:RGES}) is satisfied. 
\end{theorem}

\textbf{Proof:} The proof follows similar steps as the proof of Theorem 1 in \cite{Allan2019} (compare also the earlier results in \cite{Mueller2017}) with suitable adaptations for the data-based setting and substantial simplifications due to the considered linear system. 

 For $t \leq L-1$, each term in the cost function of problem~(\ref{MHE_nom_ini}) can be upper bounded by the optimal total cost, which in turn by optimality can be upper bounded by the cost of the real (unknown) trajectory. A lower bound for the optimal value function can be found by exploiting Assumption \ref{ass:quadrati_costs}. Using these upper and lower bounds, the following statements are obtained for $t = L-1$ (compare \cite[Eqs. (6) and (7)]{Allan2019})
\begin{align}
|\hat{e}_0|& \leq  \sqrt{2\frac{p_2}{p_1}}   |e_0|  \oplus  \sqrt{\frac{2Lr_2}{\rho p_1}}||\mathbf{v}||_{[0,L-1]}\label{e_hat} ,\\ 
||\hat{\boldsymbol{\sigma}}^y||_{[0,L-1]} & \leq \sqrt{2\rho p_2/r_1}|e_0| \oplus    \sqrt{2L r_2/r_1}   ||\mathbf{v}||_{[0,L-1]}  \label{sigma},
\end{align}
where $\hat{e}_0 \coloneqq \hat{x}(0|L-1) - \hat{x} (0)$, $e_0 \coloneqq  \hat{x} (0) -x(0)$ and $\hat{x}(0) = \hat{x}_0$, $ x(0) = x_0$. The expression $\hat{\boldsymbol{\sigma}}^y$ denotes the optimal fitting error sequence and $\mathbf{v}$ the real (unknown) fitting error sequence. Furthermore, since ($A$,$C$) is detectable, it is e-UOSS (compare \cite[Theorem 6]{Knufer2020}). Hence we can use~(\ref{Definition_UOSS}) with $x_1 = \hat{x}(0|t)$ and $x_2 = x(0)$. By (\ref{System_Dynamics_nom_ini}) and the definition of the estimated state $\hat{x}$, for $0\leq t\leq L-1$ we have 
\begin{equation*}
\hat{x}(t)=x(t; x_1, \mathbf{u})= H_1(x^d_{[t,N-1]}) \hat{\alpha}(t).
\end{equation*}
Furthermore, regarding the real (unknown) state trajectory of system (\ref{system_def}), it follows, by applying Theorem \ref{thm:Willems} (with $y=x$) that there exists an $\alpha$ such that
\begin{equation*}
x(t)=x(t; x_2, \mathbf{u})= H_1(x^d_{[t,N-1]}) \alpha(t). 
\end{equation*}
Additionally
, we have that $h(\mathbf{x}_1, \mathbf{u}) = H_{t+1}(y^d_{[0,N-1]}) \hat{\alpha} = \tilde{\boldsymbol{y}} - \hat{\boldsymbol{\sigma}}^y $ and $h(\mathbf{x}_2, \mathbf{u}) = H_{t+1}(y^d_{[0,N-1]})\alpha = \tilde{\boldsymbol{y}}- \mathbf{v}$ (where $\mathbf{x}_1$ and $\mathbf{x}_2$ are as in Definition~\ref{Definition_UOSS_Text}), leading to
\begin{align}
|e(t)|  \leq &   c_\beta |\hat{x}(0|t) -x(0)| \lambda_\beta^t  \oplus c_{\gamma_d} ||\mathbf{v} - \boldsymbol{\hat{\sigma}}^y ||_{[0,t-1]} \nonumber \\
 \leq & c_\beta (2|e_0| \oplus 2|\hat{e}_0 |) \lambda_\beta^t \oplus 2 c_{\gamma_d} ||\mathbf{v}||_{[0,t-1]} \nonumber \\
 & \oplus 2 c_{\gamma_d} ||\boldsymbol{\hat{\sigma}}^y ||_{[0,t-1]}  , \label{eq_intermediate1}
\end{align}
where $e(t) = \hat{x}(t) - x(t)$. We employ the derived bounds in~(\ref{e_hat}) and (\ref{sigma}) and plug them into (\ref{eq_intermediate1}). Since $||\boldsymbol{\hat{\sigma}}^y||_{[0,t-1]} \leq  ||\boldsymbol{\hat{\sigma}}^y||_{[0,L-1]}$ for all $0\leq t\leq L-1$, we obtain
\begin{align}
|e(t)| \leq& c_\beta \lambda_\beta^t  2|e_0| \oplus  c_\beta \lambda_\beta^t  2  \sqrt{2 p_2/p_1}  |e_0|  \nonumber \\
& \oplus c_\beta \lambda_\beta^t  2 \sqrt{\frac{2Lr_2}{\rho p_1}} ||\mathbf{v}||_{[0,L-1]}  
 \nonumber \\
& \oplus 2 c_{\gamma_d} ||\mathbf{v}||_{[0,t-1]} \oplus  2 c_{\gamma_d} \sqrt{2\rho p_2/r_1} 
|e_0| \nonumber \\
&\oplus 2 c_{\gamma_d} \sqrt{2L r_2/r_1} ||\mathbf{v}||_{[0,L-1]}.
\end{align}
Using $ s \leq  \sqrt{2p_2/p_1} s$ and $ s \leq \sqrt{2L r_2 /r_1} s$ this expression can be simplified to
\begin{align}
|e(t)| &\leq c_\beta \lambda_\beta^t  2  \sqrt{2\frac{p_2}{p_1}}  |e_0|  \oplus c_\beta \lambda_\beta^t  2 \sqrt{\frac{2Lr_2}{\rho p_1}} ||\mathbf{v}||_{[0,L-1]}  
 \nonumber \\
& \oplus  2 c_{\gamma_d} \sqrt{2\rho p_2/r_1} 
|e_0| \oplus 2 c_{\gamma_d} \sqrt{2Lr_2/r_1} ||\mathbf{v}||_{[0,L-1]}. \label{Proposition_7}
\end{align}
Next we apply (\ref{Proposition_7}) with $ t = L-1$ and a time shift of $\tilde{k}$, which can be done by choice of the prior in (\ref{eq:Gamma_MHE}), i.e., at time $\tilde{k}$, the prior $\hat{x}(\tilde{k}-L+1)$ is used. Hence we obtain
\begin{align}
|e&(\tilde{k} + L-1)| \leq c_\beta \lambda_\beta^{L-1}  2  \sqrt{2p_2/p_1}  |e(\tilde{k})|  
 \nonumber \\
&\oplus  c_\beta \lambda_\beta^{L-1}  2 \sqrt{\frac{2Lr_2}{\rho p_1}} ||\mathbf{v}||_{[\tilde{k},\tilde{k} + L-1]}  
 \nonumber \\
& \oplus  2 c_{\gamma_d} \sqrt{2\rho p_2/r_1} 
|e(\tilde{k})| \oplus 2 c_{\gamma_d} \sqrt{2Lr_2/r_1} ||\mathbf{v}||_{[\tilde{k},\tilde{k} + L-1]}.      \label{Proposition_7_extended}
\end{align}
Fix $\lambda \in (0,1)$ and note that for all 
\begin{equation*}
L \geq \left\lceil \ln{\Bigg(\frac{\lambda}{c_\beta 2 \sqrt{2p_2/p_1}}\Bigg)} / \ln{(\lambda_\beta)}+1 \right \rceil\eqqcolon T
\end{equation*}
it follows that
\begin{equation*}
c_\beta \lambda_\beta^{L-1}  2  \sqrt{2p_2/p_1}  s  \leq \lambda s
\end{equation*}
holds for all $s \in \mathbb{R}^n$. Additionally, one can find a $\rho$ small enough such that
\begin{equation*}
2 c_{\gamma_d} \sqrt{2\rho p_2/r_1} s \leq \lambda s 
\end{equation*}
for all $s \in \mathbb{R}^n$. These arguments are now used for the first and for the third term of (\ref{Proposition_7_extended})
\begin{align}
|e(\tilde{k} + L&-1)| \leq \lambda |e(\tilde{k})| \oplus   c_\beta \lambda_\beta^{L-1}  2 \sqrt{\frac{2Lr_2}{\rho p_1}} ||\mathbf{v}||_{[\tilde{k},\tilde{k} + L-1]}  
 \nonumber \\
& \oplus   \lambda |e(\tilde{k})| \oplus 2 c_{\gamma_d} \sqrt{2Lr_2/r_1} ||\mathbf{v}||_{[\tilde{k},\tilde{k} + L-1]} \nonumber \\
 &\leq \lambda |e(\tilde{k})| \oplus c_{\gamma_e} ||\mathbf{v}||_{[\tilde{k}, \tilde{k} + L-1 ]}
 \label{Proposition_7_extended_2}
\end{align}
where 
\begin{align}
 c_{\gamma_e} \coloneqq    c_\beta   2 \sqrt{\frac{2Lr_2}{\rho p_1}} \oplus 2 c_{\gamma_d} \sqrt{2L\frac{r_2}{r_1}}  . \label{def:gamma}
\end{align}
Next, we establish upper bounds for $|e(t)|$ for $t \in \mathbb{I}_{[0, L-2]}$. By similar computations that lead to (\ref{Proposition_7}), it follows
\begin{align}
|e(&t)| \leq 2c_\beta   \sqrt{2\frac{p_2}{p_1}}  |e_0|  \oplus  2 c_\beta   \sqrt{2\frac{L r_2}{\rho p_1 }}  ||\mathbf{v}||_{[0,t-1]}  \nonumber \\
& \oplus   2 c_{\gamma_d} \sqrt{2 \rho p_2/r_1} |e_0| \oplus 2 c_{\gamma_d}              \sqrt{2L r_2/r_1}          ||\mathbf{v}||_{[0,t-1]}    \nonumber \\
 &= c_{3}  |e_0|   \oplus c_{\gamma_e}||\mathbf{v}||_{[0, t-1]}
\end{align}
for $c_{3} \coloneqq 2c_\beta   \sqrt{2p_2/p_1}  \geq\sqrt{p_2/p_1} \geq 1 > \lambda  \geq 2 c_{\gamma_d} \sqrt{2 \rho p_2/r_1}$. The remainder of the proof is to show by induction that 
\begin{align}
 |e(t &+j(L-1))|\leq \lambda^j c_{3}|e_0| \oplus c_{\gamma_e}|| \mathbf{v}||_{[j(L-1), j(L-1) + t]} \nonumber \\
 & \oplus \mathrm{max}_{i \in  \mathbb{I}_{[0, j-1]}}\big( \lambda^{j-i-1}  c_{\gamma_e}||\mathbf{v}||_{[i(L-1), (i+1)(L-1) -1]}\big)\nonumber
\end{align}
for all $j \geq 0$ and $ t \in \mathbb{I}_{[0, L-2]}$ which works analogously to the model-based case in \cite{Allan2019}. Hence, for the sake of brevity, we do not show this part here. Define 
\begin{align}
c_{1} \coloneqq \frac{c_{3}}{\lambda^{\frac{L-2}{2(L-1)}}} \quad \quad  c_{2} \coloneqq \frac{c_{\gamma_e}}{\lambda}
\end{align}   
as well as $\lambda_{1} = \lambda_{2}  \coloneqq \lambda^{\frac{1}{2(L-1)}}$. Then, for all $t \geq 0$
\begin{equation}
|x(t) - \hat{x}(t)| \leq  c_{1} |x_0 - \hat{x}_0| \lambda_{1}^t \oplus \: \bigoplus_{\tau = 0}^{t} c_{2} |v(t-\tau)| \lambda_{2}^\tau \label{eq:final_proof}
\end{equation} 
which proves that the data-based MHE is RGES. \hfill  $\blacksquare$

\begin{remark}
\label{rmk:conservatism}
The gain $c_{\gamma_e}$ defined in (\ref{def:gamma}) increases with an increasing horizon $L$. This rather counter-intuitive result is analogous to the model-based setting \cite{Mueller2017,Allan2019} and results from various conservative steps in the employed proof technique. Recently, novel proof techniques using a Lyapunov-like analyis \cite{Allan2020} or using discounted cost functions \cite{Knufer2020} have been proposed to circumvent this problem. Studying similar concepts for the proposed data-based MHE framework is subject of future work.
\end{remark}

\begin{remark}
In the data-based MHE problem (\ref{MHE_noisy}), one can allow for more general cost functions, which are not necessarily quadratic. In this case, one needs to consider \mbox{Assumptions 2 - 5} of \cite{Allan2019} (which are satisfied globally for the quadratic costs considered here) for the stability analysis. Depending on whether these assumptions hold locally or globally, the resulting robust stability guarantees then hold locally or globally.
\end{remark} 

\begin{remark}
In both model-based and data-based MHE, past disturbance sequences are estimated when solving the corresponding MHE optimization problems, allowing for a similar proof technique. On the other hand, in the ``dual" problem of MPC, the model-based results (compare, e.g.,~\cite{Rawlings2017}) and the data-based robust stability proofs~\cite{Berberich2020_MPC} are quite different. The reason is that in data-based MPC, the noise in previous online output measurements (and in the offline collected data) is estimated via an additional slack variable (similar to (\ref{System_Dynamics_noisy})), which is not the case in most model-based robust MPC schemes. 
\end{remark}

\section{Numerical Example}
\label{sec:Illustrative_Example}
\begin{figure}[t!]
\begin{subfigure}[h]{1\linewidth}
	\centering
  \includegraphics[width = 1\linewidth]{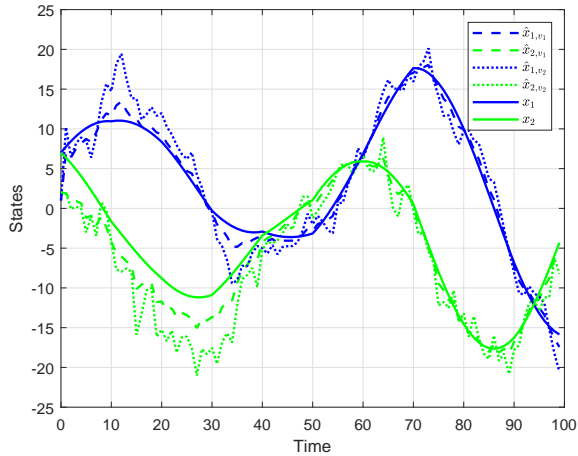}
   \caption{$R = 10 I_p$, $P =10 I_n$}
   \end{subfigure}
   \begin{subfigure}[h]{1\linewidth}
   	\centering
   \includegraphics[width = 1\linewidth]{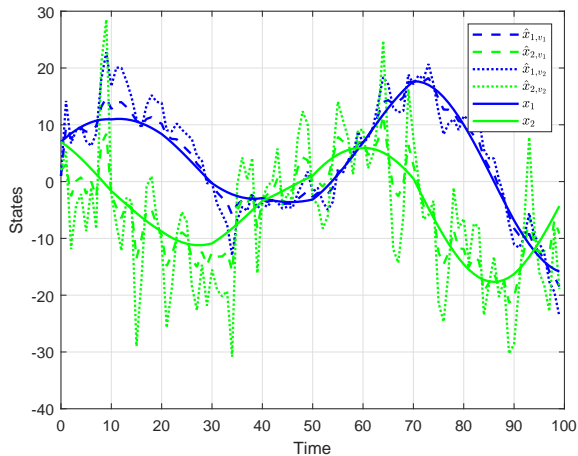}
  \caption{$R = 100 I_p$, $P =10 I_n$}
   \end{subfigure}
\caption{This figure illustrates the behavior of the data-based MHE scheme for two different choices of $R$ and $P$ in the case of measurement noise.}
\label{fig:Simulation_Results}
\end{figure}

In this section, we illustrate our data-based MHE scheme by means of a simple numerical example taken from \cite{Kwon1999}. The following LTI-DT system
\begin{subequations}
\begin{align*}
\begin{bmatrix}
x_1(k+1) \\
x_2(k+1) \\
\end{bmatrix}   &= \begin{bmatrix}
0.9950 & 0.0998 \\
-0.0998 & 0.9950 \\
\end{bmatrix} \begin{bmatrix}
x_1(k) \\
x_2(k) \\
\end{bmatrix} 
+ \begin{bmatrix}
0.1 \\ 
0.1 \\
\end{bmatrix} u(k) \\
y(k) &= \begin{bmatrix}
1 & 0 \\
\end{bmatrix}
x(k) + v(k) 
\end{align*} 
\end{subequations}
is considered. We collect offline a persistently exciting input/output trajectory of length $N = 30$, where the input is sampled from a uniform random distribution $\mathcal{U}(-5,5)$. The measurement noise is normally distributed, i.e., $v \sim \mathcal{N}(\mu, \sigma^2)$. The data-based MHE scheme is applied for two different cases: first for the case in which $\mu = 0$ and $\sigma = 2$ (named $v_1$) and, second, $\mu = 0$ and $\sigma = 6$ (named $v_2$). We choose $\rho = 1$, $L=5$, and consider $R = 10 I_p$, $P =10 I_n$ and in a second simulation $R = 100 I_p$, $P =10 I_n$. The initial condition for the system is $x_0 = \begin{bmatrix}
7& 7
\end{bmatrix}^\top$ and for the prior estimate $\hat{x}_0 = \begin{bmatrix}
1 & 2
\end{bmatrix}^\top$. The simulation results are illustrated in Figure \ref{fig:Simulation_Results}. As guaranteed by Theorem \ref{thm:RGES}, the data-based MHE is RGES. Naturally, a higher noise variance leads to worse estimates. 

Interestingly, a higher value for $R$ leads to a higher variation of $\hat{x}_2$ (and to a slightly higher variation of $\hat{x}_1$). A higher value of $R$ implies that we prefer small values of $\sigma^y$ such that the estimated output, i.e., $H_L(y^d_{[0,N-1]})\alpha(t)$, is closer to the measured, noisy output. Consequently, the estimates will be influenced substantially by the noisy output measurements, which explains the higher variations of $\hat{x}_2$ (and the slightly higher variations of $\hat{x}_1$).

Although not addressed in this simulation example (and throughout this paper), an interesting topic for future research is a detailed comparison between the proposed (direct) data-based MHE scheme to (indirect) model-based MHE schemes using system identification, similar to the analysis in the context of data-driven predictive control \cite{Pasqualetti2021}.



\section{Conclusion}
\label{sec:Conclusion}
In this paper, we introduced a data-based MHE scheme for the case in which the online output measurements are corrupted by some non-vanishing measurement noise. We showed RGES of the robust data-based MHE and illustrated its performance by means of a simulation example. Besides studying alternative data-based MHE formulations or proof techniques that potentially allow for less conservative results (compare Remark~\ref{rmk:conservatism}), our current work focuses on how bounded noise in the offline collected data influences our results.

\bibliographystyle{IEEEtran}
\bibliography{IEEEabrv,Literature_Data_Based_MHE}

\end{document}